\documentclass[aps,prd,twocolumn,superscriptaddress,showpacs]{revtex4}
\usepackage{bm}
\usepackage{amsmath}
\usepackage{rotating}
\usepackage{aas_macros}

\newcommand\lsim{\mathrel{\rlap{\lower4pt\hbox{\hskip1pt$\sim$}}
        \raise1pt\hbox{$<$}}}
\newcommand\gsim{\mathrel{\rlap{\lower4pt\hbox{\hskip1pt$\sim$}}
        \raise1pt\hbox{$>$}}}
\newcommand\propsim{\mathrel{\rlap{\lower4pt\hbox{\hskip1pt$\sim$}}
        \raise1pt\hbox{$\propto$}}}
\newcommand{\D}{\mathrm{d}}
\newcommand{\tot}{\mathrm{tot}}
\newcommand{\ret}{\mathrm{ret}}
\newcommand{\Msun}{\mathrm{M}_{\odot}}

\newcommand{\yr}{\,\mathrm{yr}}

\begin{document}
\bibliographystyle{apsrev}

\title{Distortion of Gravitational-Wave Packets Due to their Self-Gravity}

\author{Bence Kocsis}
\email{bkocsis@cfa.harvard.edu}
\affiliation{Harvard-Smithsonian Center for Astrophysics, 60 Garden Street, Cambridge, MA 02138, USA}
\affiliation{Institute of Physics, E\"otv\"os University, P\'azm\'any P. s. 1/A, 1117 Budapest, Hungary}

\author{Abraham Loeb}
\email{aloeb@cfa.harvard.edu}
\affiliation{Harvard-Smithsonian Center for Astrophysics, 60 Garden Street, Cambridge, MA 02138, USA}

\date{\today}

\begin{abstract}
When a source emits a gravity-wave (GW) pulse over a short period of time,
the leading edge of the GW signal is redshifted more than the inner
boundary of the pulse. The GW pulse is distorted by the gravitational
effect of the self-energy residing in between these shells. We illustrate
this distortion for GW pulses from the final plunge of BH binaries, leading
to the evolution of the GW profile as a function of the radial distance
from the source. The distortion depends on the total GW energy released
$\epsilon$ and the duration of the emission $\tau$, scaled by the total
binary mass $M$. The effect should be relevant in finite box simulations
where the waveforms are extracted within a radius of $\lesssim
10^{2}M$. For characteristic emission parameters at the final plunge
between binary BHs of arbitrary spins, this effect could distort the
simulated GW templates for LIGO and LISA by a fraction of $10^{-3}$.
Accounting for the wave distortion would significantly decrease the
waveform extraction errors in numerical simulations.
\end{abstract}
\maketitle

\section{Introduction}\label{s:intro}

The observation of gravitational waves (GWs) is expected to open a new
window on the universe within the following decade. First generation GW
detectors
(InLIGO \footnote{http://www.ligo.caltech.edu/},
VIRGO \footnote{http://www.virgo.infn.it/},
TAMA \footnote{http://tamago.mtk.nao.ac.jp/},
GEO \footnote{http://geo600.aei.mpg.de/})
are already operating at or close to their design sensitivity levels and the development of the next
advanced-sensitivity GW detectors
(Advanced LIGO \footnote{http://www.ligo.caltech.edu/advLIGO/},
Advanced Virgo \footnote{http://wwwcascina.virgo.infn.it/advirgo/},
LCGT \footnote{http://www.icrr.u-tokyo.ac.jp/gr/LCGT.html}) and the
space-detector
LISA \footnote{http://www.lisa-science.org/}
are well underway.  It is now increasingly important to fully understand the
precise characteristics of the GW waveforms that we expect to observe.

The most luminous GW sources are expected to be associated with mergers of BH binaries. The physical understanding of these sources has greatly improved by recent breakthroughs in numerical relativity \cite{pre05,cam06,bak06a}. It is now finally possible to simulate the merger of a BH binary, from the initial circular inspiral, through the plunge to a common surrounding horizon, to the final ringdown, as the remnant settles down to a quiescent stationary Kerr-BH. It is widely believed now that existing simulations are sufficiently precise to allow targeted searches for these waveforms in real data \cite{bak07}. In fact, it has been recently shown that the errors are not even limited by the numerical precision of the simulation ($\sim 10^{-5}$), but the GW extraction method itself entails a much larger uncertainty ($\sim 10^{-3}$) \cite{paz06}. In this paper, we  demonstrate that the self-gravity during the propagation of gravitational radiation in the zone of wave extraction of numerical simuations leads to the distortion of the waves, corresponding to similar magnitude modifications in typical cases.

\subsection{Description of the effect}\label{s:intro:effect}

\begin{figure}[htbp]
\centering{\mbox{\includegraphics[width=6cm]{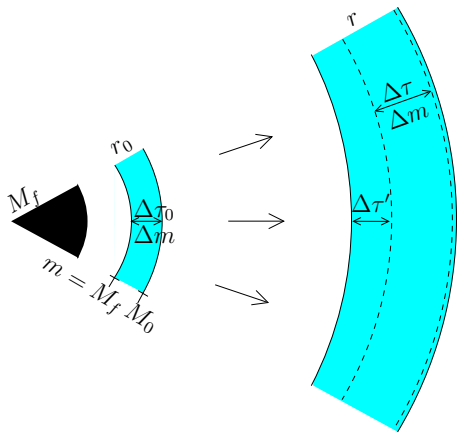}}}
 \caption{\label{fig:schematic} A sketch of the effect under consideration.
 The coalescence of two BHs in a binary of total initial mass $M_{0}$
 results in the emission of a burst of gravitational radiation which
 carries away a non-negligible $\epsilon$ fraction of $M_0$. The remnant BH mass is
 $M_{f}$. The proper temporal width of the wave-packet for a hypothetical observer
 fixed at a radial distance $r$ is $\Delta \tau$. As the packet propagates outwards, it
 (1) expands due to gravitational redshift of the initial mass $M_0$ (solid lines),
 (2) contracts due to the mean self-gravity of the radiation (dotted line, $\Delta \tau'$), and also
 (3) distorts its profile due to the self-gravity of the radiation (not shown).
 As a result the inner shells begin to catch up, and the proper time separation in
 excess of gravitational redshift from the front of the burst decreases with distance.
 Consequently, the net luminosity of the radiation burst changes with distance.}
\end{figure}

Let us imagine a compact spherically-symmetric configuration of matter
(representing the remnant) and a rapidly expanding sphere of massless
particles (representing the radiation) carrying away some of the
initial mass of the system (Fig.~\ref{fig:schematic}). First, let us assume
Newtonian gravity and spherical symmetry. In this case, the various shells
are pulled back only by the gravity of the mass {\it interior} as if it was
concentrated to a point mass at the center of the sphere, and the effect of
the outer enclosing shells exactly cancels out. Thus, the particles on the
outermost shell are always attracted by the total mass, including the mass
of the radiation, but the innermost shells experience only the gravity of
the remnant. Therefore, the gravity of the radiation implies that the
innermost shells of radiation will be continuously catching up to the outer
boundary during their journey from the source to the observer.

Do we expect an analogous effect to exist also for gravitational waves in
full general relativity? First, let us consider conventional
(i.e. non-gravitational) radiation. In analogy to the Newtonian
gravitational pull, relativistic test particles are slowed down by gravity:
the null-geodesics in a gravitational field experience the so-called {\it
Shapiro delay} \cite{sha64}, decreasing the radial coordinate velocity with
increasing gravity. Furthermore, according to Birkhoff's theorem, the
spacetime outside a spherically symmetric distribution of energy is
equivalent to the spacetime of a point-mass placed at the center of the
sphere, the Schwarzschild metric, and the spacetime inside a cavity is the
free-space Minkowski spacetime. More generally, the spherically symmetric
expansion of collisionless radiation is a known simple exact solution of
the Einstein equations, the Vaidya metric \cite{vai51a,vs60}. This solution
has exactly the same characteristics as the Newtonian example, whereby
various shells react only to the mass interior to them, i.e. they move on
world lines neglecting the exterior shells and the effect of the interior
shells is the same as if they were concentrated to a point mass at the
center.

Next, let us turn to the case of gravitational radiation. The effect of the
self-energy of gravitational radiation can be accounted for in the first
nonlinear-order approximation of the Einstein field equations by attaching
terms of order $h^2$ to the stress-energy, $T_{ij}$, considering these
terms as sources in addition to the regular radiation fields
\cite{bh64,isa68a,isa68b,mtw,tk75,tho80,efr94}. Here $h_{ij}$ is the wave
amplitude which is the correction to the background metric. If the
wavelength of the GW wave-packet is much smaller than the size of the wave
envelope, the evolution of the wave-packet is determined by the WKB
cycle-averaged effective stress-energy tensor \cite{isa68a,isa68b},
independent of the specific wave-characteristics of the radiation. In this
regime, we may treat the GW packet as an ensemble of relativistic particles
for which our previous arguments apply. In conclusion, we anticipate that
\begin{enumerate}
 \item[(i)] the wave-envelope will continuously expand due to the redshift of the
  initial mass of the binary,

 \item[(ii)] it will contract due to the self-gravity of the radiation, and

\item[(iii)] in analogy to electrodynamics, we expect that the distortion
of the wave envelope would lead to a continuous adiabatic modification in the
GW frequency.
\end{enumerate}
The purpose of this paper, is to quantify these expectations for typical BH merger waveforms using simple models and to demonstrate that this effect should be accounted for in relation to numerical simulations and observed merger waveforms.

\subsection{Related literature}\label{s:intro:literature}

To our knowledge the effect of self-gravitational distortion of GWs had not been explicitly recognized previously. We elaborate on the relation of the self-distortion effect to numerical general relativity, analytical investigations like the multipolar post-Minkowskian (MPM) and post-Newtonian (PN) theory, and the studies of the scattering of gravitational radiation in curved spacetimes.

The self-gravitational distortion of GWs is a relatively small effect on short
scales currently accessible to numerical simulations. Current
state-of-the-art simulations of binary BH mergers are restricted to the
central strong-gravity domain near the black holes, and extract gravity
waves from the boundary of this domain. The standard method of extracting
and extrapolating the waveforms to larger distances, is based on the
Regge--Wheeler--Zerilli-Moncrief perturbation formalism
\cite{rw57,zer70,mon74}. This is a linear-order representation of the
Einstein field equations and so it neglects self-energy effects of order
$h^2$. Cumulative nonlinear effects like the self-distortion effect should
lead to systematical deformations of the linear waveform extracted at
different radii, which can in principle be discovered by a rigorous
convergence test. In fact, nearly all papers on simulated merger GWs study
the convergence behaviour in some detail. However, due to computational
limitations, the extraction radius is currently restricted to $r\lsim 50
M$, and the extraction has been preformed on only a few, typically 3--4
different radii with the extrapolation done empirically based on these
radii
\cite{pre05,cam06,bak06a,bak07,bru06,bak04,clz06a,clz06b,tm07,pk07,ber07,bcp06,bak06b}. Recently,
Pazos et al. \cite{paz06} reported a systematic effect of order $10^{-3}$
(for extraction radii $r \leq 80 M$), which is much larger than the
numerical precision of the simulation $\sim 10^{-5}$. This is roughly the
same order of magnitude systematic effect that we expect for the GW
distortion for the given radii (see below). Note however, that Berti et
al. \cite{ber07} showed that the convergence behavior is also largely
sensitive to simulation assumptions. On small scales probed by these
simulations, other near-field nonlinearities might also be of equal
importance.

A precise treatment of the waveforms at large radii is possible by
analytical methods, such as the MPM-expansion introduced by Thorne
\cite{tho80} (see also \cite{tk75}) and developed further by Blanchet \&
Damour \cite{bd84,bd86,bd92}. In Ref.~\cite{tho80}, Thorne introduced the
concept of ``local-wave zone,'' which is outside the dynamical zone of wave
generation, but where nonlinear effects are still important. In this region
the propagation of gravity waves is expressed in terms of an expansion in
powers of $h_{ij}$ as an infinite sum of multipole contributions with
rapidly decreasing amplitude, which needs to be matched to the dynamical
gravitational field generated by the source. The GWs in the local wave zone
are given formally by the MPM expansion, whose terms correspond to
different powers of the gravitational coupling constant $G$. In the PN
approach, the dynamical wave generation is calculated analytically in an
infinite series in the inverse speed of light $c^{-2}$. Matching the PN and
MPM expansions in the local wave zone is a successful method for the
calculation for steady source GWs produced by relatively slow motions,
like the inspiral phase of BH mergers where the distance between the BHs is
large enough to allow an adiabatic quasicircular orbit at $r>6M$. To date,
the PN waveforms for circular binary inspirals are available to 3.5PN order
for general mass ratios \cite{bla06} (which is the highest order that is
expected to have a measurable contribution for circular inspirals by a
LISA-type detector \cite{aru06}) and 5.5PN order for extreme mass ratios
and no BH spins \cite{tts96}. To our best knownledge, the self-gravitational distortion effect has not been identified in these works. However, in this paper we show that the radius for a fixed luminosity shift or a fixed frequency shift is linearly sensitive to the energy density of the radiation. The luminosity at the inspiral phase of binaries is less than $1\%$ of the luminosity at the final plunge \cite{bcp06}. Therefore, even if it is negligible for inspirals, the self-gravitational modulation of GWs could be significant for the final plunge.

We expect the self-gravitational distortion of GWs to be consistent with
the PN/MPM expansion, and to have corresponding PN tail counterparts \cite{bla98c}. The tails of
GWs are caused by the scattering of linear waves on the spacetime curvature
generated by the total mass-energy of the source \cite{ww91}, which is
related to the Shapiro time-delay \cite{sha64} of the radiation crawling
out of the background gravity of the source. At $2.5$PN order beyond the
Newtonian quadrupole formula, GW tails scatter off the monopole field of
the remnant \cite{poi93,wis93,bs93}. Furthermore, above 3PN order, the
tails of the tails are produced by curvature scattering of the tails of the
waves themselves, associated with the cubic nonlinear interaction between
two mass monopole moments and the mass quadrupole of the source
\cite{bla98a}. In this paper we show that the modification caused by the
spherical self-gravitational distortion of GWs is to lowest order proportional to
the original waveform (i.e. without this effect) times the energy density of the waveform. Since the energy density is proportional to
the square of the amplitude, this possibly implies to lowest order a monopole-quadrupole$^2$ type interaction counterpart.

The self-gravitational distortion effect is also related to the ``memory
effects'' (or ``hereditary effects'') of gravitational radiation, since it
is a cumulative effect that depends on the full past history of the
radiation, as opposed to regular PN terms which only depend on the
instantaneous retarded fields. Other known hereditary GW effects are the
tails of GWs \cite{bd92,bla98b}, the Christodoulou effect
\cite{chr91,tho92}, and the GW recoil kick \cite{gd03}.

Finally, nonlinear effects were also examined for the interaction of plane
GWs on a free space (i.e. Minkowski) background
\cite{bon69a,as71,hs94,mc02,ser03,men03}. The nonlinear terms in the
scattering problem are found to exactly cancel up to fourth order, leaving
no self-phase-modulation effect for GWs in vacuum to this order. However,
the geometry of this case is very different from the one discussed in this
paper where a curved background is initially present. It is also unclear
whether there are self-phase modulation effects at higher nonlinear
orders. Fortunately, our approach does not face such convergence
issues, since we adopt the exact (i.e. non-perturbative) solution of the
Einstein equations in the spherical WKB approximation of an expanding
radiation shell.

The present paper aims to quantify the self-gravitational distortion
effect for recently compiled merger waveforms.
In \S~\ref{s:waveforms}, we list the main properties of the waveforms
relevant for our study.
In \S~\ref{s:spherical} we present our analysis in spherical symmetry, and derive the results for the self-gravitational distortion of signal duration and the luminosity profile.
In \S~\ref{s:discussion} we summarize the main
conclusions, and then discuss their implications. Finally, we discuss
the validity of the spherical approximation and consider the possible
effect of the anisotropy in the Appendix. We use units with ${\rm G}={\rm c}=1$
and a metric signature of $(-1,1,1,1)$.

\section{Merger waveforms}\label{s:waveforms}
To illustrate our effect, we adopt a simplified treatment for the merger GW
waveforms. Table~\ref{t:mass} lists the total radiated mass $\Delta m_{\rm
tot}$ relative to the initial total mass $M_0$ for various encounters
between compact objects found in the literature. While all of these
encounters would have a nonnegligible GW self-gravitational effect, the
inspiral--merger--ringdown events are expected to have the most prominent
event rates for interferometric GW detectors. Inspiral detection rate
estimates are between $0.3-3\yr^{-1}$ and several per day for NS/BH mergers
for inLIGO and adLIGO, respectively \cite{ngf06}; $3$ and $100\yr^{-1}$ for
stellar BH/BH inspirals in globular clusters \cite{ole06} and in galactic
nuclei \cite{pzm00} with adLIGO, respectively; $1$--$100\yr^{-1}$ and
$30\yr^{-1}$ for supermassive (SMBH) and intermediate mass BH (IMBH)
\cite{Wyithe1,pz06,mic07} and SMBH/SMBH inspirals \cite{Wyithe2,shmv05} with
LISA, respectively. The dynamic time of the encounter is proportional to
the total mass; consequently we do not consider extreme mass ratio
inspiral--mergers (EMRI), as the GW luminosity of these sources is much
smaller. Although high velocity stellar BH encounters can be very bright in
GWs, these events are expected to be rare, less than $1\yr^{-1}$ for adLIGO
or LISA \cite{koc06b}. In this analysis, we focus on equal mass
inspiral--merger--ringdown waveforms.

\begin{table}
\caption{\label{t:mass} Total radiated mass relative to the initial total
mass, $\epsilon$, for bright GW encounters. For a detailed comparison of BH
inspiral computations see Ref.~\cite{bcpz07}.}
\begin{ruledtabular}
\begin{tabular}{cccccr}
  Objects 	& Encounter\footnotemark[1]	& Spins		&  orien-     		& Refs.		& $\epsilon$\\
  && $\left(\frac{S_1}{m_1^{2}},\frac{S_2}{m_2^{2}}\right)$	&  tation\footnotemark[2]		& 		 &$[\%]$\\
  \hline
    NS--BH      & head-on	& (0.0,0.0)	&  			& \cite{lra06}	& 0.01	\\
    BH--BH      & head-on	& (0.0,0.0)	&  			& \cite{clz06c}	& 0.05	\\
    BH--BH      & head-on	& (0.1,0.1)	&   			& \cite{clz06c}	& 0.06	\\
    NS--BH      & orbiting	& (tidal\footnotemark[3],0)	&  	& \cite{lk99a,lk99b} & 0.1\\
    BH--BH      & head-on	& (0.2,0.2)	&   			& \cite{clz06c}	& 0.12	\\
    BH--BH      & head-on	& (0.0,0.0)	&  			& \cite{spe05}	& 0.13	\\
    BH--BH      & whirl		& (0.0,0.0) 	&			& \cite{pk07}	& 0.5--3\\
BH--BH      & grazing	& (0.9,0.7)     & $- \perp -$		& \cite{alc01}	& 0.9	\\
    BH--BH      & grazing	& (0.9,0.7)     & $+ \perp +$		&\cite{alc01}	& 1.0	\\
    BH--BH      & grazing	& (0.0,0.0)     & 	  		&\cite{alc01}	& 1.2	\\
    BH--BH      & inspiral	& (0.2,0.2)	& $+ \parallel +$	& \cite{bak04}	& 1.8	\\
    BH--BH      & inspiral	& (0.1,0.1)	& $- \parallel -$	& \cite{bak04}	& 2.0	\\
    BH--BH      & inspiral	& (0.2,0.2)	& $- \parallel -$	& \cite{bak04}	& 2.0	\\
    BH--BH      & inspiral	& (0.8,0.8)	& $- \parallel -$	& \cite{clz06b}	& 2.2	\\
    BH--BH      & inspiral	& (0.1,0.1)	& $+ \parallel +$	& \cite{bak04}	& 2.4	\\
    BH--BH      & inspiral	& (0.0,0.0)	&			& \cite{bak04}	& 2.5	\\
    BH--BH      & inspiral	& (0.0,0.0)	&   			& \cite{bak06a,clz06a}	& 3.2	\\
    BH--BH      & inspiral	& (0.0,0.0)	&   			& \cite{clz06b,bru06}	& 3.5	\\
    BH--BH      & inspiral	& (0.0,0.0)	&   			& \cite{bak06b,ber07}	& 3.7\footnotemark[4]	\\
    BH--BH      & inspiral	& (0.1,0.1)	& $+ \parallel +$	& \cite{bcp06}	& 5.2	\\
    BH--BH      & inspiral	& (0.8,0.8)	& general\footnotemark[5] & \cite{tm07}	& 5--6	\\
    BH--BH      & inspiral	& (0.8,0.8)	& $+ \parallel +$	& \cite{clz06b}	& 6.7	\\
    BH--BH      & r.whirl	& (0.0,0.0) 	&			& \cite{pk07}	& 15\footnotemark[6]\\
    BH--BH      & r.head-on	& (0.0,0.0) 	&			& \cite{dp92}	& 16	\\
    BH--BH      & r.head-on	& (1.0,0.0)	& $\perp$		& \cite{cl03}	& 17$\frac{\eta}{1/4}$\footnotemark[7]\\
    BH--BH      & whirl		& (0.0,0.0) 	&			& \cite{pk07}	& 24\footnotemark[8]\\
    BH--BH      & r.whirl	& (0.0,0.0) 	&			& \cite{pk07}	& 100\footnotemark[8]\\
\end{tabular}
\footnotetext[1]{Here ``head-on'' stands for a direct collision with
$v\parallel r$ initially, ``orbiting'' stands for the tidal stripping of a
NS by a BH in close orbit, ``grazing'' stands for inspiral collisions in
which the initial separation is within the final orbit in a merger,
``inspiral'' is the complete inspiral--merger--ringdown event, ``whirl''
stands for particles approaching from infinity with some impact parameter
leading to a quasi-circular whirl-type orbit before merger, ``r.whirl'' and
``r.head-on'' corresponds to relativistic initial velocities $v\approx 1$
at $r\gg M$.}  \footnotetext[2]{Here we give the sign of $S_1\cdot J$, the
relationship between $S_1$ and $S_2$, and the sign of $S_2\cdot J$, where $J$ is the orbital angular momentum.
In case of a head-on collision with spins, we give the initial direction of
the spin relative to the separation vector.}  \footnotetext[3]{NS tidally
locked} \footnotetext[4]{Ref~\cite{ber07} provides the results for
different mass ratios, $m_1/m_2=1$--$4$, and found that $\Delta m\propto
\eta^2$, where $\eta=m_1m_2/(m_1+m_2)^2\leq 1/4$.}  \footnotetext[5]{8
different choices of spin orientations} \footnotetext[6]{In case the impact
parameter is small enough to end up in a merger.}
\footnotetext[7]{Calculated for $m_1\ll m_2$. (See $\eta$ above at
\footnotemark[4].)}  \footnotetext[8]{In case the impact parameters are
fine tuned for the binary to approach the unstable circular orbit.}
\end{ruledtabular}
\end{table}

Generally, the waveforms can be expanded in multipoles \cite{tho80}
\begin{equation}
 h_{\mu\nu} = \sum_n\frac{(1+z){\cal A}^n_{\mu\nu}}{d_L} e^{i\phi_n}\label{e:h_munu}
\end{equation}
where $\phi$ is the high frequency GW phase which is related to the
instantaneous frequency through $f=\D \phi/\D t$; ${\cal A}_{\mu\nu}$ is a
slowly varying envelope describing how the instantaneous amplitude changes
over the waveform, the index $n$ labels the various polarizations and
multipoles, and $d_L$ is the luminosity distance, and $z$ is the redshift.
For binary inspiral merger waveforms, the ($l=2,m=2$) multipole
(i.e. quadrupole) dominates the waveform \cite{bcp06,bak06b,ber07}. For the
sake of simplicity, we restrict our attention to a single monochromatic
wave.

During a single GW cycle, corresponding to a time interval $f^{-1}$, the
envelope $A_{\mu\nu}$ can be regarded as constant. In the WKB
approximation, the energy carried by the radiation can be calculated as a
cycle-averaged quantity. The effective stress-energy tensor of the
radiation is \cite{isa68b,mtw}
\begin{equation}
 T^{\mu\nu}= \frac{1}{64\pi} \frac{A^2}{d_L^2} k^{\mu}
 k^{\nu},\label{e:Tmunu}
\end{equation}
where $A^2={\cal A}^{\mu\nu}{\cal A}_{\mu \nu}$ is the squared effective amplitude and $k_{\mu}=\phi_{,\mu}$ is the wave number.

Each infinitesimal volume of the wave can be attributed the mass-energy
that it carries according to Eq. (\ref{e:Tmunu}). In the spherically
symmetric approximation, the total luminosity (or more precisely the
graviton number) is \cite{isa68b}
\begin{equation}
 L = \frac{\D E}{\D r\D t} = 4\pi r^2 T^{t r}= \frac{A^2}{16}   k^t k^r \label{e:lum}
\end{equation}
With this equation it is possible to obtain the waveform $[A(t),f(t)]$ within radius $r$ from the luminosity function $L(t)$.

Based on the waveforms derived by numerical simulations (such as Fig.~25 in
Ref. \cite{bcp06}), we adopt the following simple fit to the effective
luminosity
\begin{equation}
 L(t) =
 L_0\times \left\{
  \begin{array}{cc}
    [(t_1-t)/t_1]^{-1.5}   & \text{if~}-t_0<t<0\\
    1             & \text{if~}0<t<t_1 \\
    \exp[-(t-t_1)/t_2] & \text{if~}t>t_1\\
  \end{array}
\right.\label{e:waveform}
\end{equation}
where the intervals $t<0$, $0< t<t_1$ and $t_1>t$ correspond to the late
inspiral/final orbits, the peak luminosity at the plunge, and the ringdown
phases, respectively, $-t_0$ sets the initial time of the calculated
profile, $t_1$ represents the characteristic timescale of the most
intensive part of the radiation, $t_2$ sets the ringdown decay rate, and
$L_0$ is the normalization luminosity. Numerical simulations show
\cite{bcp06} that the characteristic frequency of the radiation rapidly
increases after the inspiral and saturates at $\omega=2\pi f \sim
M_0^{-1}/2$
at $t\sim 0$, where $M_0$ is the initial mass of the source.
The characteristic number of wave cycles during the brightest phase is
$N=\Delta t/f^{-1}\sim 3$ over a time $\Delta t= 12\pi M_0$.
Note that we assume that the WKB method is applicable to the waveform,
implying that $L(t)$ does not change greatly over a single cycle. This
condition is just marginally satisfied for these waveforms.

For a simple analysis, we assume that the total luminosity crossing a
sphere at infinity $L_{\infty}(t)$ is given by Eq.~(\ref{e:waveform})
with the following parameters: $(t_0,t_1,t_2)=(100, 10, 5)M_0$, and use
$\Delta t= 12\pi M_0$ in reference to average quantities below.  Under the
WKB approximation the luminosity (\ref{e:waveform}) evolves independently
of the carrier frequency $f\sim 1/(4\pi M_0)$.  The $L_0$ normalization is
set using the total radiated mass of the system as a fraction of $M_0$
by $\epsilon=\Delta m_{\tot}/M_0$ for the inspiral-mergers (\ref{e:lum})
listed in Table~\ref{t:mass}. For pedagogical comparison purposes we
distinguish the inspiral events only based on the normalization $L_0$, and
do not consider the variations in the shape of the waveform (e.g. $\Delta
t$).

\section{Quantitative estimates in spherical symmetry}\label{s:spherical}

To give a first quantitative estimate of the magnitude of the wave-packet
distortion due to self-energy, we start by computing the propagation of
unpolarized radiation packets in the spherically symmetric Vaidya
spacetime. The possible effect of anisotropies is discussed in the
Appendix.

The Vaidya metric \cite{vai51a} in radiation coordinates $(u, r, \theta,
\phi)$ is
\begin{equation}
 \D s^2 = -\left(1-\frac{2m(u)}{r}\right)\D u^2 - 2 \D r \D u + r^2 \D\Omega,
\label{e:Vaidya}
\end{equation}
where $d\Omega = d\theta^2 + \sin^2\theta d\phi^2$. This metric is an exact
(i.e. nonperturbative) solution of Einstein's equations in spherical
symmetry in the eikonal approximation to a radial flow of unpolarized
radiation.  Here $u$ is the retarded time parameter which is constant along
the world lines of radially outgoing radiation,
and $m(u)$ describes the mass function interior to $u$.
Outside the radiation (i.e. where $m(u)$ is constant in the space-time),
the Vaidya metric (\ref{e:Vaidya}) is the Schwarzschild solution
in Eddington-Finkelstein coordinates \cite{mtw}. For our approximate waveforms
decribed in \S~\ref{s:waveforms}, $m(u)$ is $M_0$ constant outside,
it is quickly changing within a short range $0\leq u \leq\Delta u_{\tot}$ across,
and it is $M_f=M_0-\Delta m_{\tot}$ inside the radiation shell. Here, $\Delta m_{\tot}$ and
$\Delta u_{\tot}$ are set by the simulated waveforms \S~\ref{s:waveforms}.

For the metric given by Eq.~(\ref{e:Vaidya}) it is straightforward to
derive the convergence of null-geodesics describing the world lines of
radiation shells using Raychaudhuri's equation or the equation of geodesic
deviation \cite{wald}.
In either way, we find that radially outgoing shells of radiation simply follow
the world lines $u(r) = \rm constant$, implying that the $\Delta u$ coordinate difference between the shells
does not change during the propagation. The physical contraction of radiation shells can be
examined using the proper time measure between shells and the observed luminosity
profile.

\subsection{Proper time duration}\label{s:spherical:time}
First, we estimate the proper-time duration of the GW signal along the world-line of a hypothetical observer crossing the radiation shell. For simplicity, we restrict to observers at a fixed spacial coordinate $(r,\theta,\phi)$. Then we have $\D r \equiv \D\theta \equiv \D \Omega\equiv 0$
and Eq.~(\ref{e:Vaidya}) gives
\begin{equation}
 \D \tau^2 = -\D s^2 = \left(1-\frac{2m(u)}{r}\right)\D u^2
\end{equation}
leading to $\D \tau = \sqrt{1-2m(u)/r}\D u$. Therefore, $\D u$ can be interpreted as the infinitesimal proper time difference between two
radiation shells at fixed radius approaching infinity. Thus, we adopt the notation $\D u \equiv \D \tau_{\infty}$, and similarly
$\Delta u \equiv \Delta \tau_{\infty}$ for integrated quantities. Finally, let us define $m_i=m(u_i)$ for $i\in \{1,2\}$ and $\Delta m=|m_2-m_1|$. We set $(m_1,m_2,\Delta m_{\tot})=(M_0,M_f,\epsilon M_0)$ when referring to the total GW signal duration.

Integrating between two arbitrary shells of radiation $u_2$ and $u_1$ gives
\begin{equation}\label{e:Dtau}
 \Delta \tau = \int^{\tau_2}_{\tau_1}\D \tau = \int^{u_2}_{u_1} \left( 1- \frac{2m(u)}{r} \right)^{1/2} \D u.
\end{equation}
After expanding $m(u)$ in a Taylor-series, the integrand becomes
\begin{equation}
  \left(1- \frac{2m_1}{r} - \frac{2u\langle m'\rangle}{r} - \frac{u^2\langle m''\rangle}{r}+\dots\right)^{1/2}.
\end{equation}
Substituting in Eq.~(\ref{e:Dtau}), and using $\langle m'\rangle=-\Delta m/\Delta \tau_{\infty}$ and $\langle m''\rangle=0$ to first order, we get
\begin{equation}\label{e:Dtau_1}
 \Delta \tau(m_1,m_2,r) =  \frac{r\Delta \tau_{\infty}}{3|m_2-m_1|}
\left.\left(1-\frac{2m}{r}\right)^{3/2}\right|^{m_2}_{m_1}.
\end{equation}
Setting $m_1=M_0$ and expanding in terms of $\Delta m$, we get
\begin{equation}
 \frac{\Delta \tau}{\Delta \tau_{\infty}} =  \sqrt{1- \frac{2M_0}{r}}\left( 1 + \frac{1}{2}\frac{\Delta m}{r-2M_0} + {\cal O}(\Delta m^2)\right).
\label{e:Dtau_2a}
\end{equation}
The leading order term can be indentified as the gravitational redshift for
constant mass $M_0$, the second term describes the correction due to the
radiation mass. If we expand also in terms of powers of $1/r$, the relative
change in the proper time duration of the signal becomes
\begin{equation}
 \frac{\Delta \tau-\Delta \tau_{\infty}}{\Delta \tau_{\infty}} =- \left(M_0 - \frac{\Delta m}{2}\right)\frac{1}{r}  + {\cal O}(r^{-2},\Delta m^2).
\label{e:Dtau_2}
\end{equation}

Equations~(\ref{e:Dtau_1}--\ref{e:Dtau_2}) describe the
self-gravitational distortion between radiation shells in terms
of proper time along world lines of $r=\rm const$. One can notice that to
leading order this is simply the gravitational redshift for the average
enclosed mass between the shells. However, since $\Delta m$ changes along the
wave packet non-trivally for a fixed radius as a function of time, the modification of the profile is
generally {\em not} self-similar, leading to the distortion of the luminosity
profile as a function of radius.

To make this point clearer, we correct for the average distortion of the
signal and calculate the residual distortion of the signal. Let us define
\begin{equation}
 \Delta \tau' = (1+z)\Delta \tau - \Delta \tau_{\infty}
\label{e:Dtau'}
\end{equation}
where $(1+z)$ is a time-independent constant representing the ``average gravitational redshift''
at a given $r$, given by
\begin{equation}\label{e:1+z}
  1+z = \frac{1}{\sqrt{1-\frac{2\langle m \rangle}{r}}},
%1 - \left(M_0 - \frac{\Delta m_{\tot}}{2}\right)\frac{1}{r},
\end{equation}
$\langle m\rangle\equiv (m_1+m_2)/2$ is the average mass, and to leading order
\begin{equation}\label{e:z}
  z \approx \frac{\langle m\rangle}{r}= \frac{M_0}{r} - \frac{\Delta m}{2r}.
\end{equation}

Now let us take an arbitrary radiation shell enclosing mass $\Delta m$ relative to the shell enclosing mass $M_0-0.5\Delta m_{\tot}$, i.e.
we set $m_1=M_0-0.5\Delta m_{\tot}$ and $m_2=m_1-\Delta m$ in eq.~(\ref{e:Dtau_1}). After correcting for the average gravitational redshift using Eqs.~(\ref{e:Dtau'}) and (\ref{e:1+z}), the residual relative distortion to leading order is
\begin{equation}
 \frac{\Delta \tau'}{\Delta \tau_{\infty}} =- \frac{\Delta m}{2 r}  + {\cal O}(r^{-2},\Delta m^2).
\label{e:Dtau_3}
\end{equation}
\begin{figure}[htbp]
\centering{ \mbox{\includegraphics[width=8.5cm]{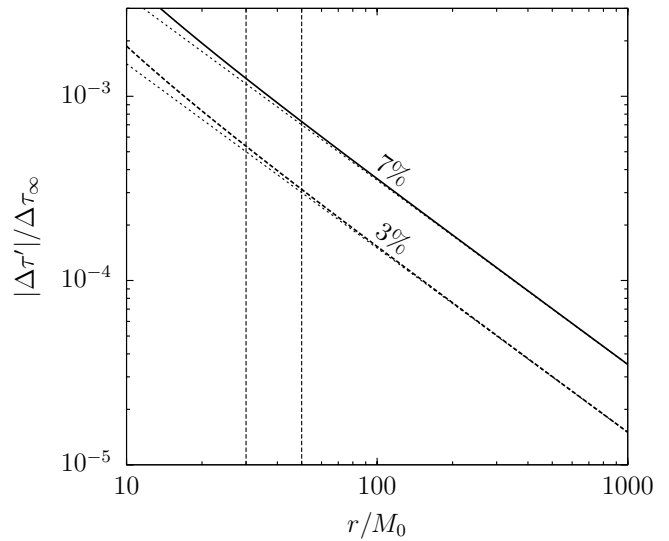}}}
\caption{\label{fig:Dtau_r} The residual self-gravitational distortion of shells after correcting for the bulk gravitational redshift. The thick curves show the change in the proper time duration of the signal at radial distance $r$ from the source between shells enclosing $7\%$ {\it (top)} or $3\%$ {\it (bottom)} of the total mass, dotted lines correspond to the leading order term given by Eq.~(\ref{e:Dtau_3}).  The vertical lines highlight the typical radii used in numerical simulations for GW extraction.}
\end{figure}
Figure~\ref{fig:Dtau_r} shows the residual distortion using the exact formula
(Eq.~(\ref{e:Dtau'}), thick lines) and the leading order contribution
(given by Eq.~(\ref{e:Dtau_3}), dotted lines).
At the typical radius used by numerical simulations
for waveform extraction, $r=50\Msun$, the primary bulk waveform distortion
changes the signal duration by $M_0/r\sim 2\%$, and the secondary relative
waveform distorsion between the front and the back of the signal is
$\epsilon M_0/(2r)\sim 7\times 10^{-4}$ for a typical BH inspiral--merger
with high spins $\epsilon=7\%$ (see \S~{\ref{s:waveforms}}). The figure also shows
that the higher order effects beyond $1/r$ lead to an uncertainty of order
$10^{-3}$--$10^{-4}$ for $r=(30$--$50)M_0$.

\subsection{Luminosity Profile}\label{s:spherical:luminosity}
Since Eqs.~(\ref{e:Dtau_1}) and (\ref{e:Dtau_2}) are applicable to two arbitrary shells of radiation, we can use them to compute the evolution of an arbitrary initial radiation profile, whereas the luminosity is simply $L=\Delta m/\Delta \tau$, the total mass-energy crossing a sphere at radius $r$ within proper time $\Delta \tau$.

The profile at infinity is given by $m(u)$ in radiation
coordinates, or $\Delta \tau_{\infty}(m)$, the proper time a shell enclosing mass
$m$ arrives at $r=R$ where $R\rightarrow \infty$, relative to the outermost
shell of radiation \footnote{Throughout this paper we assume that the
metric at $r\rightarrow\infty$ is the Minkowski metric, neglecting
cosmological effects. This is justified by the orders of magnitudes or
radii where the radiation shell distortion is relevant $r<10^{3}M_{0}$ (see
Fig.~\ref{fig:Dtau_r}) is much less than cosmological scales $r\gsim
10^{18}M_{0}$.}. Here $\Delta \tau_{\infty}(m)$ can be any monotonically
decreasing function, for which the luminosity at $R$ in Eq.~(\ref{e:lum})
is
\begin{equation}
 L_{\infty}(\tau)= -\left[\frac{\D \Delta \tau_{\infty}(m)}{\D m}\right]^{-1},
 \label{e:L_0}
\end{equation}
where the minus sign originates from our definition of $m$: the shell
labeled by the largest value of $m$ arrives the earliest.  The luminosity
profile can also be obtained as the function of time, $L_{\infty}(\tau)$,
using the relationship $\Delta \tau(m)$. Conversely, for given
$L_{\infty}(\tau)$, we can compute $\Delta \tau_{\infty}(m)$ using
Eq.~(\ref{e:L_0}). The luminosity at some other distance $r$ can be
obtained similarly if given $\Delta \tau(m,r)$, the arrival time of mass
$m$ relative to the outermost shell at distance $r$. This function is given
by Eq.~(\ref{e:Dtau_1}), substituting the waveform $\Delta \tau_{\infty}(m)$ for $\Delta \tau_{\infty}$,
and $(m_1,m_2)=(M_0,m)$. The
luminosity profile at $r$ is then
\begin{align}
 L_r(m)=&-\left[\frac{\partial \Delta \tau(m,r)}{\partial m}\right]^{-1}
%\nonumber\\
=\left(1-\frac{2m}{r}\right)^{-1/2}L_{\infty}(m)
\label{e:L_r(m)}
\end{align}
Therefore, the modification of the profile in Bondi radiative
coordinates $(m,r)$ the profile is distorted self-similarly.
However, in terms of the observer proper time variable,
$\Delta \tau(m) = \int_{0}^{m}L_r(m)^{-1}\D m$, $L_r(\tau)$,
the modification to the profile will {\em not} be self-similar:
\begin{align}
 L_r(\tau)
=\left(1-\frac{2\int_{0}^{\tau}L_r(\tau')\D \tau'}{r}\right)^{-1/2}L_{\infty}\left(\int_{0}^{\tau}L_r(\tau')\D \tau'\right)
\label{e:L_r(m)2}
\end{align}
Equation~(\ref{e:L_r(m)2}) relates the luminosity profile as a
function of proper time at radius $r$, $L_{r}(\tau)$, to the profile
at infinity, $L_{\infty}(\tau)$. Comparing Eqs.~(\ref{e:L_r(m)})
and (\ref{e:L_r(m)2}) shows the advantage of Bondi type
radiative coordiantes as opposed to proper time.

\begin{figure}[htbp]
\centering
\mbox{\includegraphics[width=8.5cm]{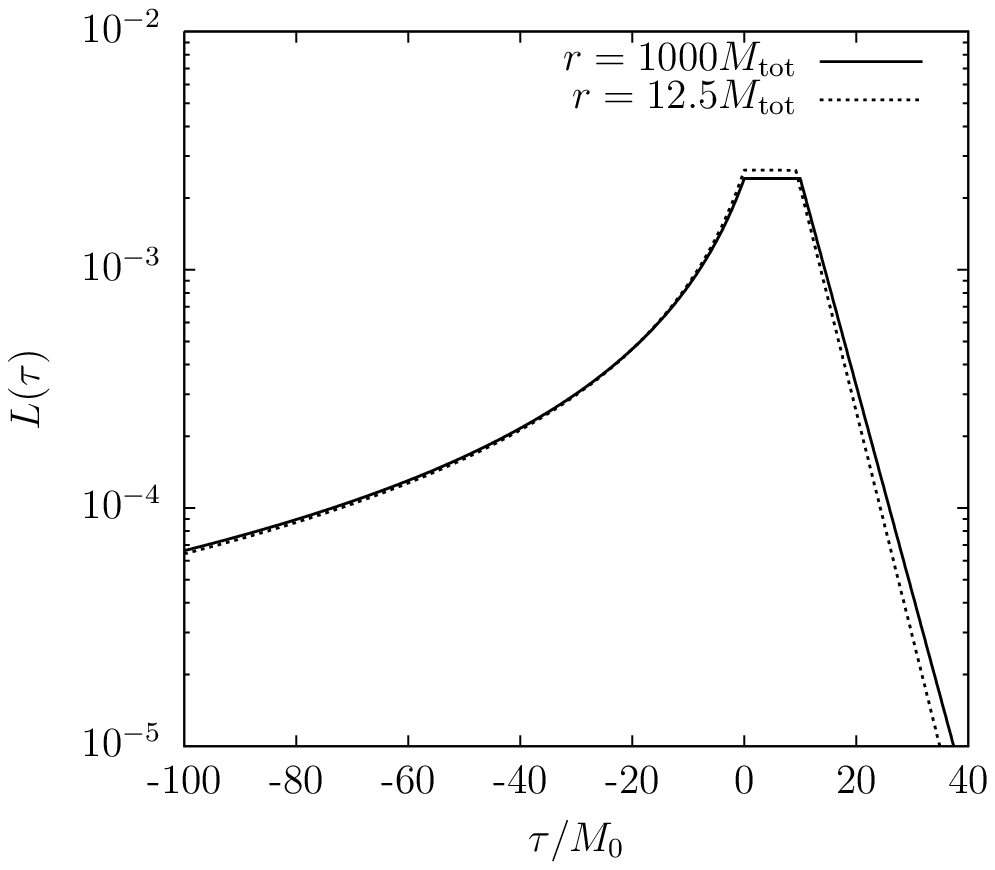}}\\
\mbox{\includegraphics[width=8.5cm]{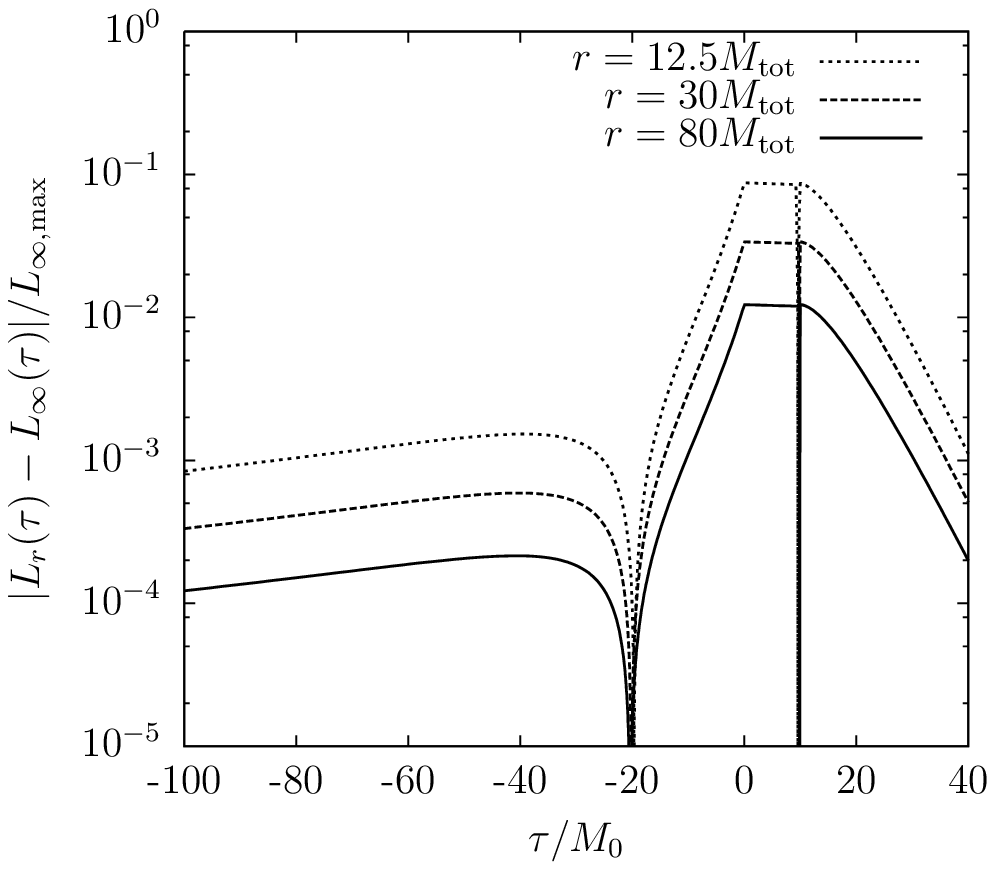}}
\caption{\label{fig:Lum_t} Our fits to the GW luminosity profile for binary
BH inspiral merger simulations as a function of observer
proper time $\tau$ and the evolution of the profile at various distances,
$r$. The profile parameters are given in \S~\ref{s:waveforms} and $\epsilon=7\%$.
{\it Top:} The absolute profile (in units of ${\rm c}^5/{\rm G}$) is
shown for two extremes, a nearby distance ($r=12.5M_{0}$) and far-away
distance ($r=10^3M_{0}$).  {\it Bottom:} The difference between the GW
luminosity profiles at infinity (i.e. $r=10^3M_{0}$) and three cases of
smaller $r$, in units of peak luminosity at infinity. The peaks of the
profiles are set to $\tau=0$. The main effect responsible for the
differences seen in this figure is the bulk gravitational redshift.}
\end{figure}
Figure~\ref{fig:Lum_t} plots $L_r(\tau)$ for our fit to the
luminosity profile at infinity of merging binary BHs $L_{\infty}(\tau)$ with $\epsilon=7\%$
(see \S~\ref{s:waveforms}). The top panel shows the absolute profile while
the bottom panel shows the difference between the profile at some radius
$r$ and the profile at infinity, such that the peak of the profiles are at
$\tau=0$. The bottom panel is useful to visualize the characteristic
evolution of the profile.

\begin{figure}[htbp]
\centering \mbox{\includegraphics[width=8.5cm]{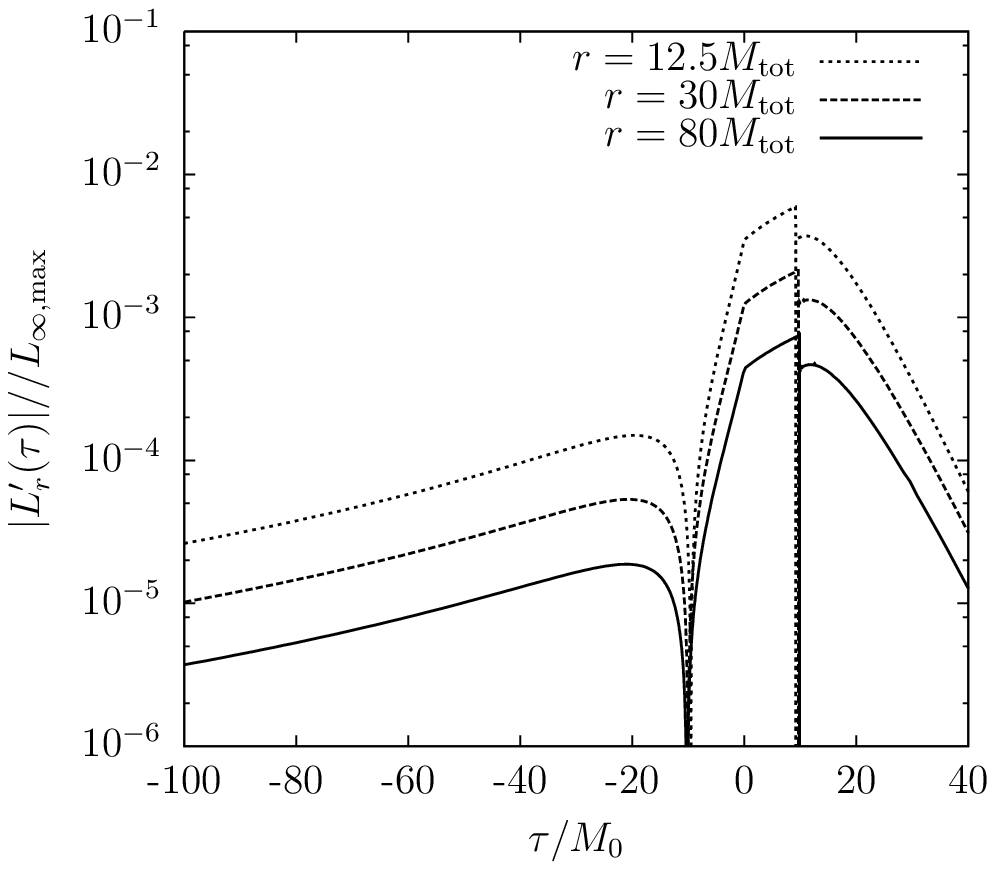}}\\
\mbox{\includegraphics[width=8.5cm]{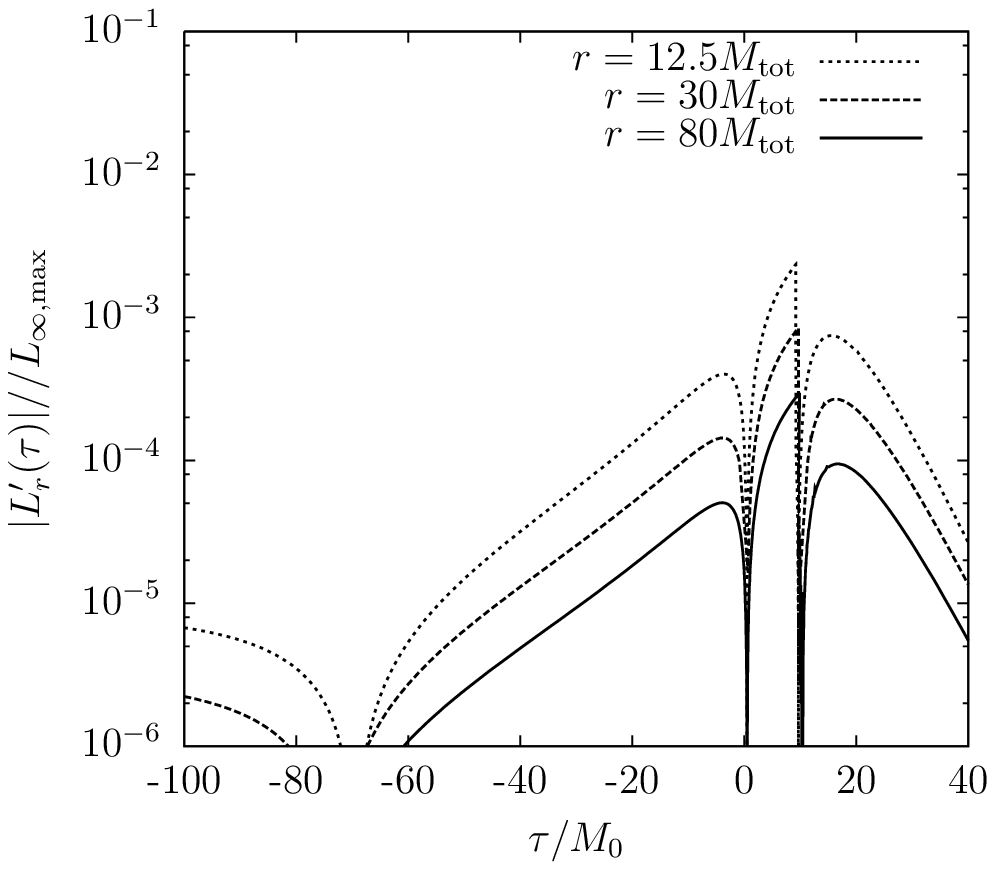}}
\caption{\label{fig:Lum_self_t} The residual self-gravitational distortion to the luminosity profiles after accounting for the avarage gravitational redshift, $z=M_{0}/r$ {\it (top)} or $(M_{0}-0.5\Delta m_{\tot})/r$ {\it (bottom)}, respectively. Other parameters are the same as in Fig. 3.}
\end{figure}
In \S~\ref{s:spherical:time}, we have identified the two main effects
responsible for the convergence rate of the waveform to be the
gravitational redshift corresponding to the average mass and the
self-gravitational effect. Indeed, the differences visible in
Figure~\ref{fig:Lum_t} are primarily due to the former. However, correcting
for only the average gravitational redshift at each radius $r$ leaves a
nonnegligible systematic error with respect to the true signal. To see
this, we substitute $\Delta \tau_{\infty}/(1+z)$ given by Eq.~(\ref{e:1+z}) into
Eqs.~(\ref{e:L_0}) and (\ref{e:L_r(m)}), and refer to the corresponding luminosity as
the {\it average gravitational redshifted luminosity profile}, $L_r^{z}(\tau)$.
After subtracting from the true profile for each $\tau$, the residual luminosity distortion is
\begin{equation}
 L'_r(\tau) = L_r(\tau) - L_r^{z}(\tau),
\end{equation}
where we set the reference time again to $\tau=0$ for the peak of the
luminosity profiles.

Figure~\ref{fig:Lum_self_t} shows the residual {\it
self-gravitational distortion} $L'_r(\tau)$ for various radii in
units of the peak luminosity at infinity, $L_{\infty,\max}$. Naturally, the
definition of the ``average gravitational redshift,'' $z$, used for
defining $L'_r(\tau)$ makes a difference in the result. The top panel uses only the initial binary mass
$\langle m\rangle=M_{0}/r$ in Eq.~(\ref{e:z}) totally neglecting the gravity of the radiation shell,
while the bottom panel has $\langle m\rangle=(M_{0}-0.5\Delta m_{\tot})/r$, i.e. the redshift is
chosen to account also for the average
gravity of the radiation shell. In the later case we find a much quicker
convergence for the waveform peak at increasing radii, but the former
choice is more suitable for the early parts of the waveform corresponding
to the late inspiral waveform. A comparison of Fig.~\ref{fig:Lum_t} and
\ref{fig:Lum_self_t} shows that the self-gravitational distortion is
roughly an order of magnitude smaller than the effect of the average
gravitational redshift.

\subsection{Self-Gravitational Coordinate Effects}
\label{s:spherical:coordinate}

In the previous sections we have derived the time duration and the luminosity profile of the radiation shell as it propagates radially outward from the source. We have assumed that the GW profile at each fixed arial radius $r$ is parameterized by the proper time $\tau$ of a hypothetical observer fixed at that radius, in particular the luminosity $L_{r}(\tau)$ was the total mass-energy crossing a sphere at radius $r$ within infinitesimal proper time $\D \tau$. Therefore, the adopted time-coordinate $\tau$ corresponds to a synchronous gauge at each radius. Since physical observables depend precisely on proper measures, these coordinates allow a simple interpretation of the convergence characteristics of the GW profile at large radii.

Other choices of coordinates would have introduced additional artificial distortion effects making the convergence characteristics of the waveforms much different. Consider for instance the ``natural'' coordinate system $(t,r,\theta,\phi)$ in the spherically symmetric case that is chosen to be Schwarzschild both before and after the GW burst has arrived with masses $M_0$ and $M_f=M_0-\Delta m_{\tot}$, respectively, and which changes smoothly in between these regions. An example of such a coordinate system can be derived from the Vaidya metric Eq.~(\ref{e:Vaidya}) with the implicit transformation $t\equiv u+r+2m(u)\ln [r-2m(u)]$ where $m(u)$ describes the mass function interior to $u$ (which is constant along the outgoing radiation world lines). Indeed, everywhere in the spacetime where $\D m/\D u=0$, these coordinates yield a Schwarzschild metric, and the Vaidya metric in radiation coordinates (\ref{e:Vaidya}) is then simply the Schwarzschild solution in Eddington-Finkelstein coordinates \cite{mtw} in these regions. This map covers all relevant parts of the spacetime including the GW zone. The world-lines of radiation shells can be shown to follow
\begin{equation}\label{e:v_r}
 \frac{\D r}{\D t} = 1-\frac{2m_u}{r},
\end{equation}
where the second term is called the Shapiro-time delay \cite{sha64} for a
particle crawling out of the gravitational potential of mass $m_u$ interior
to it. After integration, we find that to first order the temporal
separation of two radiation shells enclosing mass $\Delta m$ at radius $r$
evolves to leading order as $\Delta t(r)=\Delta t(R_0) - 2\Delta m \ln
(r/R_0)$, where $R_0$ is an arbitrary initial radius. In these coordinates,
the signal duration contracts uniformly in exponential distance
intervals. Even though the metric is asymptotically Minkowski (where $t$
approaches $\tau$ for $r\gg M_0$), the resulting profile evolution is
fundamentally different from $\Delta \tau(r)$ given by
Eq.~(\ref{e:Dtau_2})!

The appearance of the logarithmic radial dependence of the waveform was
first realized by Fock \cite{fock}. This effect is specific to the harmonic
coordinates and can be avoided if changing to Bondi type radiative
coordinates \cite{iw68,mad70}. Blanchet \& Sch\"afer \cite{bs93} have shown
that a similar logarithmic dependence of the GW tail leads to a
tail-induced amplitude and phase shift (typically of order $10^{-7}$) for
stationary sources. In contrast, the logarithmic radial dependence of the
wave contraction for merger waveforms can be significant for GW merger
simulations. Between $r=(20$--$40) M_{0}$, the waveform contracts in
$\Delta t$ by a fraction of $5\times 10^{-3}$, which is just of the order
of the current wave extraction precision \cite{paz06,bak06b,bcp06}. This
apparent logarithmic contraction effects can be avoided if one changes to
the proper time variable as we have done in the previous sections. The
remaining $\Delta \tau(r)$ evolution is however a physical effect.

Both the logarithmic $\Delta t(r)$ contraction and the physical $\Delta
\tau(r)$ evolution (in particular the contribution denoted by $\Delta
\tau'(r)$ above) are consequences of higher order radiation effects in the
Einstein equations beyond the scope of first order methods such as the
Regge--Wheeler--Zerilli-Moncrief perturbation method used for extrapolating
the numerical waveforms to infinity. Therefore these effects cause the
extrapolated waveforms to be different when extracting GWs from numerical
simulations at various radii by standard methods using no
self-gravitational interaction. For a related recent analysis see Ref.~\cite{lm07}.

\section{Discussion}\label{s:discussion}
\subsection{Summary}\label{s:summary}

We considered the self-gravitational effect of gravitational radiation on
the propagation of GWs from a compact source. We adopted simple
approximations for the geometry of the radiation, by considering spherical
symmetry on scales comparable to the radial width of the radiation
packet. This approximation appears adequate for the quadrupolar $(l=2,m=2)$
radiation pattern around binary BH sources in numerical simulations
\cite{bcp06,bak06b,ber07}. Nevertheless, we use the Appendix to examine the
maximal effects of anisotropy in the opposite (exaggerated) extreme, when
the outgoing radiation is concentrated into a compact region. We find that
irrespective of the level of anisotropy, the gravitational radiation is
distorted under the influence of its own gravity as it propagates. Contrary
to the standard gravitational redshift, which is a uniform shift of the
waveform, the self-gravitational effect depends on the intensity and is
predominant only for the most intensive bursts of radiation causing a
non-uniform distortion of the waveform. The self-gravitational distortion
depends on distance to leading order as $\Delta m/r$, and is therefore
relevant on scales $r_{\rm sg}/M_{0}\lsim \epsilon/\delta$ where $\epsilon$
is the radiation efficiency and $\delta$ is the desired calculation
accuracy. For BH binary mergers simulations $\epsilon\sim 7\%$ and
$\delta\sim 10^{-5}$ implying that $r_{\rm sg}\lsim 7\times
10^{3}M_{0}$. If the GWs are extracted within this region, the
self-gravitational distortion should be taken into account.

\subsection{Testing the Effect with Numerical Simulations}\label{s:discussion:simulations}

Numerical simulations based on the full set of Einstein equations for
binary BH inspirals have not yet reported evidence for the waveform
distortion effect considered here although they have shown that the
waveforms do not converge within a fractional accuracy of $\delta \sim
10^{-3}$ \cite{bak06b,bcp06,paz06}. This is because the simulations are
restricted to a limited volume, typically of radii $\sim (80$--$850)M$,
while the extracted waveforms are typically compared between $r\sim
20$--$50M$. For a radiation mass $\Delta m/M_0\sim 7\%$, the primary effect
is a shift of the waveform due to a logarithmic Shapiro time delay of the
remnant, a uniform gravitational redshift, and the self-gravitational
effect. We have shown that the logarithmic Shapiro delay does not show up
if using proper measures to describe the waveform, and the uniform
gravitational redshift is accounted for in the linear wave propagation
models. However, the residual self-gravitational effect in the GW
luminosity has a characteristic profile that has to be subtracted when
extrapolating the extracted waveform. The peak of the effective luminosity
distortion reaches $2\times 10^{-3}$ and $5\times 10^{-4}$ at $r=30$ and
$50M_{0}$, respectively.

Present-day numerical relativity simulations should already be capable of directly measuring the relevance of our effect by artificially amplifying the gravitational radiation found at the extraction radius $r\sim 20M$, and starting the simulation with these amplified initial conditions. For example, for a total GW energy $\Delta m/M_0=30\%$, the effective luminosity distortion between $r=20M$ between $20$--$50M$ is several percent, which is well within simulation and extraction errors. For consistency, the simulation should confirm that the total energy content of the radiation does not change. We also expect the initial ringdown frequency (corresponding to the most energetic shell) to be smaller than the final ringdown frequency.

In order to avoid errors caused by the self-gravitational distortion
effect up to the desired numerical precision $\delta\sim 10^{-5}$, the
waveform extraction radius should be chosen to be $r_{\rm sg}\gsim 7\times
10^{3}M_{0}$. Alternatively, if the waveforms are extracted at smaller
radii, the waveforms should be converted to Bondi type radiative
coordinates and then extrapolated with the scaling $1/r$.

\subsection{Observational Implications}\label{s:Discussion:Implications}
The self-gravitational waveform distortion is important for future observations of BH binary mergers.

\begin{enumerate}
\item The waveform distortion is expected to be resolvable for the LISA
instrument with respect to simulated waveforms for total BH masses of
$(10^4$--$10^9)\Msun$. The total signal to noise ratio of merger waveforms
is $10^4$ for LISA observing $z_c\sim 1$ \cite{bak07}. The distortion
effect modifies the waveform amplitude and frequency by $\sim
(10^{-3}$--$10^{-4})$
for numerical waveforms extracted between $r=(30$--$80)M_{0}$.

\item The distortion involves a systematic modification of the waveform
which needs to be accounted for in order to interpret observed merger
waveforms and improve the estimation uncertainty of physical parameters
beyond the uncertainty of the preceding inspiral signal. The signal to
noise ratio of the final BH merger waveform is an order of magnitude larger
than for the inspiral, implying that the merger waveform has a potential to
greatly reduce parameter estimation errors. Note that the relative accuracy
using only the inspiral signal with LISA is expected to be
$10^{-3}$--$10^{-5}$ \cite{lh06,lh07} for estimating the component masses,
which is smaller than the distortion effect.

\item This effect is different from the uncertainties caused by
gravitational lensing \cite{2hh05}, in that it is only an issue concerning
the convergence properties of numerical simulations. Lensing causes an
error of several percent on the inferred luminosity distance (due to the
unresolved matter along the line of sight), and lensing errors increase
with the source distance. In contrast, the self-gravitational effect is of
order 0.1-0.01 percent for numerical simulations if the waveforms are
extracted at $30M$--$80M$ and dies off quickly as $1/r$. The wave
distortion effect is of order $10^{-20}$ relative to the waveform amplitude
for typical astrophysical scales. Therefore, the wave-contraction effect
does not provide any additional physical parameters for observations.
\end{enumerate}

\acknowledgments

We thank George Rybicky, Irwin Shapiro and Kip Thorne for enlightening
discussions. BK acknowledges support from a Smithsonian Astrophysical
Observatory Predoctoral Fellowship and from NKTH \"Oveges J\'ozsef
Fellowship.

\appendix
\section{Anisotropy}\label{s:anisotropy}
%Bence: the split between the text and the appendix is just fine.
%We just need to make sure that the reader will not be confused about
%the logarithmic (coordinate-related) factors.

Our analysis considered only perfectly spherically-symmetric
configurations.  The approach was motivated by the quasi-spherical
radiation patterns found around binary BH sources in numerical simulations
\cite{bcp06,bak06b,ber07}. In this Appendix, we would like to examine the
sensitivity of our basic results to deviations from sphericity.  To gauge
whether there is any such sensitivity, we analyse the most extreme case in
which the outgoing radiation is concentrated into a highly compact region.

But first let us define more precisely what we assumed so far.  The
derivation presented in \S~\ref{s:spherical} requires that the radiation
field is ``initially locally spherically symmetric'', so that it is
initially described by the Schwarzschild metric locally within some narrow
solid angle $\Delta\theta\lsim \Delta r/r$ before the GW arrives, where
$\Delta r\sim c\Delta t$ is the radial width of the wave-packet along its
propagation direction. But since the radiation propagating through this
solid angle is in no causal contact with the radiation field expanding
towards other directions, it cannot distinguish the actual spacetime from a
spherically symmetric one. Note that the outermost shells of radiation
expanding along different directions are always causally disconnected by
definition, and the interior shells of radiation can only be affected by
the outer shells within $\Delta \theta$. Since we examine the distortion effect
on large distances compared to the width of the
burst $r\gg \Delta r$, spherical symmetry must only be required within a
very narrow angle. This simple set of considerations implies that if
high-order multipoles have a vanishing contribution at large distances, the
results derived in \S~\ref{s:spherical} are applicable very generally for
short bursts of radiation, $\Delta r\ll r$. Indeed, numerical simulations
confirmed that the dominant contribution to the wave amplitude is given by
the ($l=2,m=2$) multipole and higher order terms are suppressed by more
than a factor of magnitude (see references in \S~\ref{s:waveforms}).  In
the remainder of this Appendix we demonstrate the validity of this simple
conclusion through explicit calculations.

We consider three variations on a toy model to estimate the effect of
anisotropy. We start with the simplest model and refine this model by
adding more details and complexity in the successive models. In each case,
we discuss general implications for the model under consideration.  In all
models we consider the extreme opposite regime to spherical symmetry,
namely that the radiation is maximally clumped into two outgoing BHs $L$
(leading) and $T$ (trailing) of masses $m_{L}$ and $m_{T}$, representing
the leading and trailing edges of the radiation, respectively. We assume
that $L$ and $T$ are moving in the same direction on light-like world
lines, so that $T$ lies in the causal past of $L$, but $L$ is outside the
causal past of $T$ throughout their propagation. We assume that there is
also a remnant Schwarzschild BH $R$ with mass $m_{R}$. The instantaneous
radial position coordinate of $R$, $T$, and $L$ at time $t$ are $0$,
$r_{T}(t)$, and $r_L(t)$. We are interested in obtaining the world lines of
BHs $T$ and $L$ to see how the coordinate separation $\Delta r(t)=
r_L(t)-r_T(t)$ decreases with time as compared to the spherically symmetric
result. In our first model we neglect the remnant
$R$ (setting $m_R=0$), assume that $L$ moves with constant velocity $v_L$
in free space, and calculate the trajectory of $T$ in the spacetime created
by $L$. Subsequently, we will generalize $L$ to move on a more general
world line with a slowly changing velocity $v_L(t)$. Finally, we can turn
on the remnant $R$ in addition to $L$, and include the retardation effect
when calculating the relative motion of $T$.

We note that the spacetime of BHs moving at the speed of light have been
calculated previously in Ref.~\cite{hs94}, which found that BHs moving in
the same direction do not interact. However, Ref.~\cite{hs94} assumed that
the BHs move in free space and consequently adopted $v=1$ for their
velocity. In contrast, the BHs $T$ and $L$ travel on null-geodesics in the
perturbed spacetime which is initially the Schwarzschild spacetime. This
difference gives rise to a non-trivial interaction between the BHs $T$ and
$L$.

We compare our results to the spherical case, using the $(t,r)$ coordinate system defined by Scwarzschild coordinates
before and after the radiation shells as described in \S~\ref{s:spherical:coordinate}.

\subsection{No remnant $m_R=0$, constant $v_L$ velocity}\label{s:anistropy:a}

We start by assuming that $L$ is a BH with constant velocity $v_L<1$ in
free space, and wish to calculate the world line of $T$ in this
background. Here, we assume that no remnant is present, and that $L$ and
$T$ move along the same spatial direction, which we denote by $x$. Thus it
is sufficient to restrict our attention to the two dimensions $(t,x)$ of
the spacetime.

Let us start by deriving the metric. In the coordinate system $(t',x')$
comoving with $L$, the metric is the Schwarzschild metric $\D s^2 =
-(1-\phi')\D t'^2 + (1-\phi')^{-1}\D x'^2$, where
$\phi'=2m'_{L}/|x'|$. Here $x'\equiv 0$ corresponds to the BH $L$ for all
$t'$, and $m'_{L}= m_L/\gamma$ is the rest mass of $L$, where $m_L$ is the
energy carried by $L$ in the original $(t,x)$ coordinates and
$\gamma=1/\sqrt{1-v_L^2}$ is the Lorentz factor. To derive the metric in
the $(t,x)$ coordinate system, we apply the diffeomorphism
$(t,x)=\gamma(t'+v_L x',v_L t'+x')$, i.e. a global Lorentz transformation,
\begin{eqnarray}
 \D s^{2}&=& -\frac{(1-\phi')^2-v_L^2}{(1-v_L^2)(1-\phi')}\D t^2 +
 \frac{1-v_L^2(1-\phi')^2}{(1-v_L^2)(1-\phi')}\D x^2 \nonumber\\ &&-
 \frac{2v_L\phi'(2-\phi')}{(1-v_L^2)(1-\phi')}\D t \D x\label{e:ds1a}
\end{eqnarray}
which can be rearranged as
\begin{eqnarray}
\D s^{2} &=& \frac{[v_L+(1-\phi')]\D t - [1+v_L(1-\phi')]\D
x}{\sqrt{(1-v_L^2)(1-\phi')}}\nonumber\\ && \times \frac{[v_L-(1-\phi')]\D
t - [1-v_L(1-\phi')]\D x}{\sqrt{(1-v_L^2)(1-\phi')}}.\label{e:ds1b}
\end{eqnarray}
Here $\phi'$ is to be expressed as the function of the new coordinates
$(t,x)$, i.e. $\phi' = \gamma^{-2}\phi$ where $\phi=2 m_L/|\Delta x|$,
$\Delta x=x_L-x$, and $x_L=v_Lt $ is the instantaneous position of the
singularity.

The $x_T(t)$ null-geodesics describing the world line of $T$ can be
obtained by setting $\D s^2 = 0$. Equation~(\ref{e:ds1b}) shows that there
are two solutions
\begin{align}
 [v_L+(1-\gamma^{-2}\phi)]\D t - [1+v_L(1-\gamma^{-2}\phi)]\D x_T &=
  0,\label{e:geodesic1a}\\ [v_L-(1-\gamma^{-2}\phi)]\D t -
  [1-v_L(1-\gamma^{-2}\phi)]\D x_T &= 0.\label{e:geodesic1b}
\end{align}
These differential equations can also be obtained more simply by finding
the null-geodesics in the comoving coordinates $(t',x')$ first, and
changing to the $(t,x)$ coordinates only in the resulting equation. The
null geodesics in the comoving coordinates are simply $\D x'_T/\D t' = \pm
|1-\phi'|$ (see Eq.~\ref{e:v_r}), and the Lorentz
boost coordinate transformation of this differential equation leads
instantly to (\ref{e:geodesic1a}-\ref{e:geodesic1b}). Therefore, the two
solutions (\ref{e:geodesic1a}-\ref{e:geodesic1b}) describe the null
geodesics approaching or receding the moving BH, respectively.

We would like to find the solution for the $T$ test particle approaching
the source $L$ from behind, namely Eq.~(\ref{e:geodesic1a}) for an initial
condition $x_T<x_L$. This first-order differential equation can be solved
analytically by a linear substitution. The coordinate velocity $v_T=\D
x_T(t)/\D t$ monotonously decreases from $1$ to $v_L$ as the event horizon
at $x_{L{\rm hor}}(t)=v_Lt-2\gamma^{-2}m_L$ is approached. In particular if
$v_L=1$, i.e. the source $L$ has the speed of light in the free-space
background, then the trailing test particle $T$ will not be delayed at all,
$v_T(t)=1$ for all $t$. However, if $v_L\ll 1$ then $T$ is considerably
affected by the Shapiro delay near the horizon of $L$.

In concluding the description of this model, let us summarize how the
clumpy case compares to the spherically symmetric case of expanding
radiation shells. First recall that in the spherically symmetric case, $L$
has no effect on $T$ throughout the dynamics regardless of $v_L$ or $\Delta
r$. In the clumpy case, the gravity of $L$ delays the motion of $T$. The
magnitude of this delay is significant only if both of two conditions are
violated: (a) $v_L\approx 1$ and (b) $\Delta x= x_L - x_T \gg
2\gamma^{-2}m_L$. What are the ``typical numbers'' for these quantities?
Eq.~(\ref{e:v_r}) implies that $v_L=1-2M_0/x$ (which is also true in the
clumpy case, see \S~\ref{s:anistropy:c}), implying that $\gamma^{-2}\sim 4
M_0/r$, and for binary mergers $\Delta x \sim 12 \pi M_0$, $m_L < M_0-M_f
\lsim 0.06 M_0$, we find that the two cases are equivalent to (a) $x\gg
2M_0$ and (b) $x \gg 0.2 \Delta m \gsim 0.01 M_0$. Quite clearly, these
conditions will {\it not} be violated for distances outside the dynamical
regime of strong gravity e.g. $x\gsim R_0 =30 M_0$.  Thus, we expect only
very minor modifications relative to the spherically symmetric case, even
in the most clumpy case. For a quantitative estimate we need to integrate
these modifications over the relevant distances which we describe next.

\subsection{No remnant, slowly changing $v_L(t)$}\label{s:anistropy:b}

Next we consider a slowly changing source velocity $v_L(t)$ for the BH $L$,
continue to neglect a remnant $R$, and calculate the motion of $T$ in this
spacetime. Since $L$ is assumed to move on a light-like world-line, it is
not effected by $R$ and only responds to the background created prior to
the production of the bursts. Thus we assume $L$ moves on the null-geodesic
of the background as described by (\ref{e:v_r}) with $v_L=1-2m_T/x_L$.

If $v_L$ is slowly changing, we can consider $v_L$ to be constant during
short time intervals with infinitesimal jumps on their boundaries. We can
then find the corresponding world line segments of $T$ by solving the
differential equation (\ref{e:ds1b}) and matching the boundary conditions
of the successive segments by requiring continuity. In the limit that the
length of the constant time intervals approaches zero, the world line
$x_T(t)$ at every instant is given by Eq.~(\ref{e:ds1b}) with $v_L$ now
denoting the instantaneous velocity, and $\phi'$ referring to the
instantaneous value of the potential: $\phi'=\gamma^{-2}\phi$, where
$\phi=2m_L/|\Delta x|$ with $\Delta x = x_L-x$ and $x_L=\int v_L(t)\D
t$. Thus,
\begin{equation}
 \frac{\D x_T}{\D t} = \frac{v_L +
 1-\gamma^{-2}\phi}{1+v_L(1-\gamma^{-2}\phi)} = 1 -
 \frac{2m_L(1-v_L)^{2}}{\Delta x-2m_Lv_L(1-v_L)}.\label{e:dx_T/dt_b}
\end{equation}
Substituting $v_L=1-2m_T/x_L$, the distance between the clumps of radiation
satisfies
\begin{equation}
 \frac{\D \Delta x}{\D t}=\frac{2m_T}{x_L}\left(-1 + \frac{4m_L m_T}{x_L
 \Delta x - 4m_Lm_T\left(1-2m_T x_L^{-1}\right)}\right).
\label{e:dDx/dt}
\end{equation}
Note that $x_L(t)$ can be used to express the width of the packet $\Delta
x$ as a function of $x_L$.  To simplify the result, we use $q_L=m_L/m_T$,
set the units to the Schwarzschild radius $2m_T=1$, and express
(\ref{e:dDx/dt}) in terms of the logarithmic distance variable,
$y=\ln(x_L-1)$. Then
\begin{equation}
 \frac{\D \Delta x}{\D y}=-1 + \frac{q_L}{x_L \Delta x -
 q_L\left(1-x_L^{-1}\right)}
\label{e:dDx/dx}
\end{equation}
which to first order in $1/x_L$ becomes
\begin{equation}
 \frac{\D \Delta x}{\D y}= -1 + \frac{q_L}{x_L\Delta x}.\label{e:dDx/dx_b}
\end{equation}
Equation~(\ref{e:dDx/dx_b}) shows that to leading order, the wave-packet
packet contracts linearly in terms of the logarithmic distance variable
$x$. In the spherically-symmetric case, the inner shell $T$ is not
influenced by $L$ and so $\D x_T/\D t=1$ instead of
Eq.~(\ref{e:dx_T/dt_b}), leading to $\D \Delta x/\D y = -1$. Therefore the
distortion of wave-packets in the maximally clumpy case is the same as in
the spherically symmetric case to leading order. The difference arises in
the next order given by the second term in (\ref{e:dDx/dx_b}) describing
how the gravity of $L$ Shapiro-delays the motion of $T$. This is typically
of order ($4\times 6\%)/(12\pi)\times x_L^{-1}\sim10^{-4}$ for $x_L\sim
30M_0$ and gets exponentially smaller for exponentially larger distances.

\subsection{Remnant included, changing $v_L$}\label{s:anistropy:c}

The previous model assumed no remnant (i.e. $m_R=0$), and postulated that
$T$ propagated in the spacetime of $L$ by assuming that the spacetime at
$T$ was the spacetime of $L$ {\it at the same instant}. Here we consider a
nonzero $m_R$ and account for the retardation of the effect of $L$ as
percieved by $T$. We assume a slowly changing velocity and that the
gravitational perturbations are sufficiently small to allow simple
superposition to leading order. We follow a simplified approach with the
following essential assumptions:
\begin{enumerate}

\item The initial condition is a spherically symmetric Schwarzschild
spacetime centered at $x= 0$.  \item $L$ moves on a null-geodesic $x_L(t)$
of the initial background metric of $R$ and $T$, i.e. in a Schwarzschild
metric centered at $x=0$ for all $t$ and for a mass
$m_{R}+m_{T}$. The world-line of $L$ follows (\ref{e:v_r}) accordingly.
\item $T$ moves on a null-geodesic $x_T(t)$ of the background metric of $R$
and $L$, which we assume to be a simple superposition
$g_{ij}=\eta_{ij}+\delta g^{R}_{ij}+\delta g^{L,\ret}_{ij}$. Here $\delta
g^{N}_{ij}= g^{N}_{ij} - \eta_{ij}$ for a given metric, $g^{N}_{ij}$, where
$\eta_{ij}$ is the Minkowski metric, $g^{R}_{ij}$ is the metric of the
remnant i.e. the Schwarzschild metric centered at $x=0$
for all $t$ with mass $m_{R}$. The metric $g^{L}_{ij}$ is the stationary
boosted Schwarzschild metric (\ref{e:ds1a}--\ref{e:ds1b}) with mass
$m_L$, velocity $v_L$, centered at $x_L$. The label $^{\ret}$ stands for
retardation, which we describe next separately.

\item We account for retardation by setting $\Delta t_{\ret}=t-t_{\ret}$ to
be the light-travel time from $L$ to $T$.  For this we compute the inward
propagating null-geodesics from $L$ to $T$ [i.e. between positions
$(t_{\ret},r_L(t_{\ret}))$ and $(t,r_T(t))$], based on the initial
background metric of $R$ and $T$ (i.e. neglecting the gravity of $L$).
\end{enumerate}

To find the retardation time, we note that the inward propagating null
geodesics satisfies $\D x/\D t =-[1-2(m_R+m_T)/x]$. Since this is exactly
the time-reversed world line of $L$, we get $x_T(t)= x_L(t-2\Delta
t_{\ret})$. Integrating $\D t/\D x$ for the world line of $L$ between
$x_L(t-2\Delta t_{\ret})$ and $x_L(t)$,
\begin{equation}
 2\Delta t_{\ret} = \int_{x_T(t)}^{x_L(t)} \frac{x_L}{x_L-2m_R-2m_T}\D x_L,
\end{equation}
from which
\begin{equation}
 \Delta t_{\ret}= \frac{x_L-x_T}{2} + (m_R+m_T)\ln\frac{x_L-2m_R-2m_T}{x_T
 - 2m_R-2m_T}.\label{e:dt_ret}
\end{equation}
The distance where $T$ percieves $L$ is $x_L^{\ret}(t)=x_L(t-\Delta
t_{\ret})$ and the separation is $\Delta x_{\ret}\equiv
x_L^{\ret}-x_T$. Substituting $x_T$ as $x_T=x_L-\Delta x$ and
$x_T=x_L^{\ret}-\Delta x_{\ret}$ into (\ref{e:dt_ret}), we can find $\Delta
x_{\ret}$ for given $\Delta x$ and $x_L$. To first nonvanishing order in
$1/x_L$,
\begin{equation}
\Delta x_{\ret} = \frac{\Delta x}{2}-\frac{(m_R+m_T)\Delta x^2}{4
x_L^2}.\label{e:dx_ret}
\end{equation}
The leading order term corresponds to the propagation at the speed of
light in free space, $v_L=v_T=1$. Note that the correction is proportional
to $x_L^{-2}$, which is extremely small for the physical cases beyond the
strong field zone. Finally, we define the retarded position and velocity
$x_L^{\ret}=x_L-\Delta x_{\ret}$, and $v_L^{\ret}=1-2(m_R+m_T)/x_L^{\ret}$,
which can be written in terms of $x_L$ and $\Delta x$ using
Eq. (\ref{e:dx_ret}).

The metric contribution $\delta g^{L,\ret}_{ij}$ of $L$ at $T$ at time $t$,
is the boosted Schwarzschild metric (\ref{e:ds1a},\ref{e:ds1b}) with
instantaneous velocity $v_L^{\ret}$, a singularity at $x_L^{\ret}$, and
distance $\Delta x_{\ret}$.

Now we can redo the derivation presented in \S~\ref{s:anistropy:b} to
find the motion of $T$, using the modified spacetime $g_{ij}$ given
above. Again we find two solutions for $\D s^2=0$ representing the ingoing
and outgoing radiation. Expanding the outgoing solution in a series in
$x_L^{-1}$, we find
\begin{equation}
 \left.\frac{\D x_T}{\D t}\right|_{\rm clumpy}= 1 - \frac{2q_R}{x_L} -
 \left[q_R\Delta x+\frac{2 q_L(1+q_R^2)}{\Delta
 x}\right]\frac{1}{x_L^2},\label{e:dxT/dt}
\end{equation}
where $q_i=m_i/(m_R+m_T)$ for $i\in \{R,T,L\}$ and distance units are
chosen to be the Schwarzschild radius $2(m_R+m_T)\equiv 1$. In order to get
the instantaneous shell width $\Delta x$ as a function of logarithmic
distance $y$, we can redo the manipulations of
(\ref{e:dDx/dt}--\ref{e:dDx/dx_b}) for the result (\ref{e:dxT/dt}). To
first order in $1/x_L$,
\begin{equation}
 \left.\frac{\D \Delta x}{\D y}\right|_{\rm clumpy}= -q_T + \left[q_R\Delta
 x + \frac{2 q_L(1+q_R^2)}{\Delta x}\right]\frac{1}{x_L}.\label{e:dDx/dx_c}
\end{equation}
In the limit of no remnant $q_R=0$, we almost recover the solution derived
previously in Eq.~(\ref{e:dDx/dx_b}). There is a factor 2 difference, which
is the direct consequence of the retardation of the percieved distance
$\Delta x$, which had been neglected in Eq.~(\ref{e:dDx/dx_b}).

Equation (\ref{e:dDx/dx_c}) should be contrasted to the expansion of two
spherically symmetric shells $m_T$ and $m_L$ in the presence of a remnant
$m_R$. We may expand the corresponding spherical solution of
\S~\ref{s:spherical:coordinate} in a series in $1/x_L$ to first order:
\begin{equation}
\left. \frac{\D \Delta x}{\D y}\right|_{\rm spherical}= -q_T +
 \frac{q_R\Delta x}{x_L}.\label{e:dDx/dx_cs}
\end{equation}
The first two terms in Eqs. (\ref{e:dDx/dx_c}) and (\ref{e:dDx/dx_cs}) are
identical. The correction describing the ``Shapiro delay'' of contraction
in the clumpy case due to the gravity of $L$ is $2q_L(1+q_R^2)/(x_L \Delta
x)$. Substituting typical physical values $q_R=97\%$, $q_T=3\%$, $q_R=3\%$,
and $\Delta x_0 \sim 6\pi \sim {x_L}_0$ in units of Schwarzschild radii, we
see that the correction is of order $10^{-4}$ initially, and becomes
exponentially smaller at exponentially larger distances. In summary, even
in the most extreme case of clumpiness, the radiation packet propagates to
very high precision according to the spherically symmetric description.

\bibliography{ms2}
\end{document}